\documentclass[fleqn,twoside]{article}
\usepackage{testo}

\usepackage{graphicx}
\usepackage[figuresright]{rotating}

\newcommand{\AmS}{{\protect\the\textfont2
  A\kern-.1667em\lower.5ex\hbox{M}\kern-.125emS}}
\newcommand{\bb}{b \overline{b}}

\title{Experimental Results on Heavy Quarks}

\author{A. Stocchi\address[MCSD]{Laboratoire de l'Acc\'elerateur Lin\'eaire, \\ 
        IN2P3-CNRS et Universit\'e de Paris-Sud, BP34, F-91898 Orsay Cedex, France}}

\begin{document}

\begin{abstract}
This paper reviews the results presented at the 31$^{st}$ ICHEP on Heavy Quarks, with 
emphasis on those related to the determination of the Unitarity Triangle parameters.
\vspace{1pc}
\end{abstract}

\maketitle

\section{Introduction}

\begin{figure}[htb!]
\includegraphics[width=77mm]{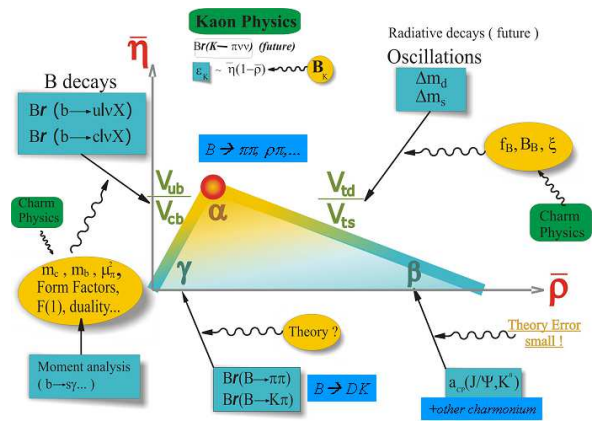}
\caption{The unitarity triangle.}
\label{fig:triangle}
\end{figure}

Accurate studies of the production and decays of beauty 
and charm hadrons are exploiting a unique laboratory for testing 
the Standard Model in the fermion sector, for studying QCD 
in the non-perturbative regime and for searching for New Physics 
through virtual processes. The first two items are the main subjects 
of this paper while the latter is discussed in \cite{nir}.\\
In the Standard Model, weak interactions among quarks are encoded
in a 3 $\times$ 3 unitary matrix: the CKM matrix. The existence of 
this matrix conveys the fact that quarks are a linear combination 
of mass eigenstates \cite{Cabibbo,km}. \\
The CKM matrix can be parametrized in terms of four free 
parameters. These parameters can be measured in several physics processes.\\
In a frequently used parametrization these parameters are named: $\lambda$, A, 
$\bar{\rho}$ and $\bar{\eta}$
\footnote{ $ \bar{\rho} = \rho ( 1-\frac{\lambda^2}{2} ) ~~~;~~~ \bar{\eta} = 
\eta ( 1-\frac{\lambda^2}{2} )$\cite{blo}.}.
The Standard Model predicts relations between the different processes 
which depend upon these parameters; CP violation is accommodated in the CKM
matrix and its existence is related to $\bar{\eta} \neq 0$.
The unitarity of the CKM matrix can be visualized as a triangle in the
$\bar{\rho}-\bar{\eta}$ plane. Several quantities, depending upon $\bar{\rho}$
and $\bar{\eta}$ can be measured and they must define compatible values for
the two parameters, if the Standard Model is the correct description of these
phenomena. Extensions of the Standard Model can provide different predictions
for the position of the upper vertex of the triangle, given by the 
$\bar{\rho}$ and $\bar{\eta}$ coordinates.\\
The most precise determination of these parameters is obtained using 
B decays and $B^0-\bar{B}^0$ oscillations.\\
Many additional measurements on B and D mesons properties (masses, 
branching fractions, lifetimes...)
are necessary to constrain the Heavy Quark theories (Operator Product Expansion (OPE) /
Heavy Quark Effective Theory (HQET) / Lattice QCD (LQCD)) to allow for precise extraction 
of the CKM parameters.  \\
Figure \ref{fig:triangle} shows ``pictorially'' the unitarity triangle
and the different measurements contributing to the determination of
its parameters.

\begin{table*}[htb]
\caption{Summary of the main characteristics of the different facilities performing $b$-physics studies.}
\label{table:bphys}
\newcommand{\m}{\hphantom{$-$}}
\newcommand{\cc}[1]{\multicolumn{1}{c}{#1}}
\begin{tabular}{@{}lllll}
 Experiments & Number of $\bb$ events &       Environment            &      Characteristics       & Status\\ \hline
 LEP   & $\sim 1M $ per expt. &       Z$^0$ decays           &  back-to-back 45 GeV b-jets   & Stopped\\
       &                  & ($\sigma_{\bb} \sim~$ 6nb)   &    all B hadrons produced     \\ 
 SLD   & $\sim 0.1M$          &       Z$^0$ decays           &  back-to-back 45 GeV b-jets & Stopped  \\
       &                  &    &    all B hadrons produced     \\
       &                  &                              &        (beam polarized)         \\  
 CLEO  & $\sim  9M $        &  $\Upsilon(4S)$ decays       &    mesons produced at rest  &  Running \\
       &                 & ($\sigma_{\bb} \sim~$ 1.2nb) &       ($B^0_d$ and $B^+$) & {(lower energies)}\\
 BaBar & $\sim  90M  $      &  $\Upsilon(4S)$ decays       &    Asymmetric B factory      &  Running \\
       &                 &  &       ($B^0_d$ and $B^+$)      \\
 Belle & $\sim  90M  $      &  $\Upsilon(4S)$ decays       &    Asymmetric B factory      &  Running \\
       &                 &  &       ($B^0_d$ and $B^+$)       \\
 CDF   & $\sim$~\rm{several ~M} & $p \overline{p}$ collider    & events triggered with & Running \\
       &                        &    $\sqrt s$ = 1.8 TeV                          & leptons or offset tracks & 
(Run II) \\
       &                        &           &    all B hadrons produced     \\  
\hline
\end{tabular}\\[2pt]
\end{table*}

In the first part of this paper we present the new results on the beauty and 
charm meson spectroscopy and lifetimes. The second part summarises 
the new results obtained in rare B decays. Part of these results, especially
those concerning the new determination of sin 2$\beta$ are described in \cite{babartalk},\cite{belletalk}.
We, then, review the results on the CKM matrix elements: $V_{cb}$ and $V_{ub}$ through B decays and 
$V_{td}$ and $V_{ts}$ using
$B^0-\bar{B}^0$ oscillations. We finally show how these measurements 
constrain the Standard Model in the fermion sector, through the determination of the 
unitarity triangle parameters.\\

B physics is studied at several facilities, which are schematically summarised in Table 
\ref{table:bphys}.\\
For D physics, at the $\Upsilon(4S)$ charm particles are produced 
in the continuum allowing B-factories to obtain
charm physics results. Charm particles are also produced in photon and hadron 
production. FOCUS experiment (E831) (the successor of E687) is designed to study charm particles 
produced by $\simeq$200 GeV photons using a fixed target spectrometer. SELEX experiment uses, instead, the 600 GeV 
Fermilab Hyperon beam (which in fact has equal fluxes of $\pi^-$ and $\Sigma^-$).

\section{Spectroscopy}
\subsection{B Spectroscopy}
\begin{figure}[htb]
\includegraphics[width=80mm]{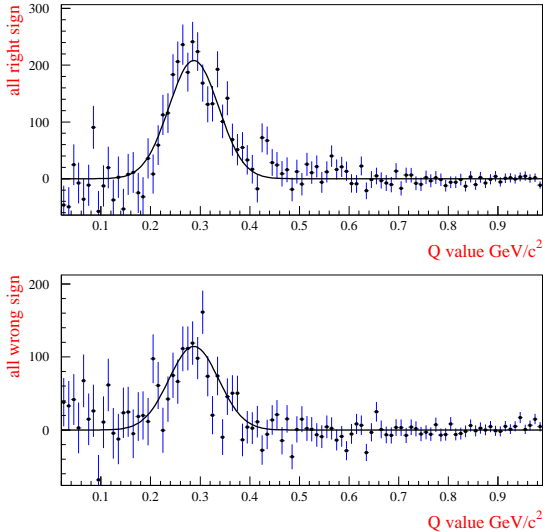}
\caption{$B^{**}_{u,d}$ analysis. The plot, from the DELPHI Coll., shows the distribution 
of $Q = [m(B\pi)-m(B)$] fitted with a single Gaussian. Results of this fit are given 
in equation \ref{eq:bssresults}. All the results are preliminary.}
\label{fig:bssdelphi}
\end{figure}

New results were presented by the DELPHI Coll. on orbitally excited B mesons (L=1, $B^{**}_{u,d,s}$) 
and excited B hadrons ($\Sigma_b^{(*)}$) . \\
\underline{$B^{**}_{u,d}$ mesons.} \\
HQET treats heavy quarks as static colour sources
and the light degrees of freedom are decoupled from the heavy quark spin. The orbitally excited states can be 
grouped into doublets of $j_q$ ($j_q = s_q + l$ ;
$s_q$ is the spin of the light quark and $l$ is its angular momentum relative to the heavy quark).
The states with $j_q$=1/2(3/2) have a broad(narrow) width, respectively.\\
Previous results were obtained by DELPHI, OPAL 
and ALEPH \cite{oldbss}. In these analyses it was not possible to separate 
the various contributions to the $B^{**}$ signals and the more plausible 
hypothesis was that both narrow and broad states contributed.
In the new results, from DELPHI, the dependence from the Monte Carlo background modelling, which was a 
critical point in the old analyses, is reduced by using purer samples and 
by fitting the background contribution directly on data.\\
The $B^{**}_{u,d}$ mesons are reconstructed, inclusively, by combining the 4-momentum 
of the B system with a charged pion having a trajectory compatible
with the primary vertex position. 
The distribution of the mass difference $Q = [m(B\pi)-m(B)$] is shown 
in Figure \ref{fig:bssdelphi}. The fit is compatible with a single Gaussian 
distribution of width corresponding to the experimental resolution, which suggests a low 
mass splitting between the narrow states. The results are:
\begin{eqnarray}
\label{eq:bssresults}
Q = 298 \pm 4 \pm 12 MeV ~;~ \sigma(Q) = 47 \pm 3 \pm 5 MeV  &\nonumber \\ 
\frac{\sigma(B^{**}_{u,d})}{\sigma_b} 
                  = (9.8 \pm 0.7 \pm 1.2) \% (\rm{narrow~states~only})
\end{eqnarray}
The data also suggest (improvement of the $\chi^2$ of the fit) the presence of a broad 
component, situated 100 MeV above the fitted narrow states component, with a width 
of $\Gamma \simeq 250 MeV$. \\
\noindent
\underline{$B^{**}_{s}$ mesons and $\Sigma_b^{(*)}$ baryons} \\
Signals from $B^{**}_s$ mesons can be obtained by replacing the $\pi$ 
candidate by an identified charged kaon. Evidence for the narrow $B^{**}_s$
mesons was found by OPAL and in a preliminary DELPHI analysis \cite{oldbssstrange}.
The new DELPHI analysis does not confirm this result and sets a limit 
on the production rate of narrow $B^{**}_s$ states:
\begin{eqnarray}
\label{eq:bsssresults}
\frac{\sigma(B^{**}_{s})}{\sigma_b} < 1.5 \% ~~\rm{at~95~\%~C.L.} \nonumber \\ 
~~~~~~~~~~~~~~ \quad \quad \quad \quad ~~(\rm{{narrow~states~only}})
\end{eqnarray}

Excited b-baryons states are the $\Sigma_b(I=1,S=1/2)$ and 
$\Sigma^*_b(I=1,S=3/2)$ in which the light diquark (ud) system has a spin and
an isospin equal to one. These baryons are expected to 
cascade into $\Lambda_b^0 \pi$. The $\Lambda_b^0$ candidate is inclusively reconstructed, 
as in previous analyses.\\ 
The new DELPHI analysis does not confirm an old preliminary evidence
\cite{oldsigmab} by setting a limit on the production rate:
\begin{equation}
\label{eq:sigmaresults}
~~~~~\frac{\sigma(\Sigma_{b}^{(*)})}{\sigma_b} ~< 1.5 \% ~~~~~~~~\rm{at~95~\%~C.L.} 
\end{equation}                
 
The Tevatron (Run II) is in a good position to obtain signals for these states, in future.

\subsection{Charmed Baryon Spectroscopy}
22 charmed baryon states were found sofar, implying a rich spectroscopy.

\begin{figure}[htb]
\includegraphics[width=80mm,height=75mm]{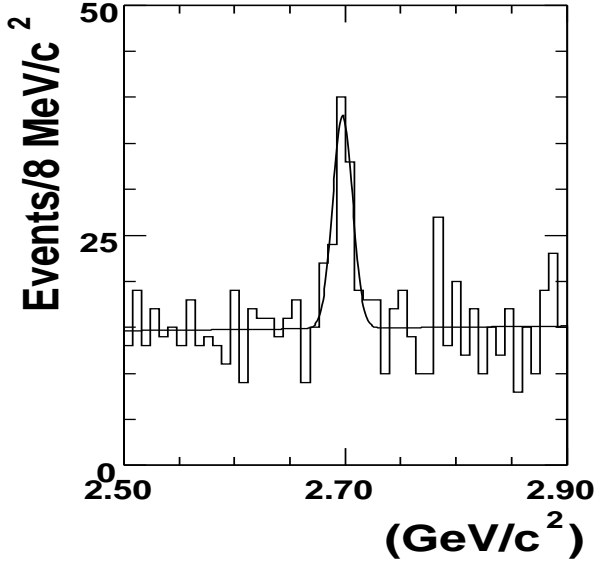}
\caption{The invariant mass spectrum of $\Omega_c$ candidates as obtained 
by the FOCUS Coll. using $\Omega^- \pi^+$ and $\Xi K \pi \pi$ decay modes. 
The result is preliminary.}
\label{fig:omegacfocus}
\end{figure}

New results were presented concerning the mass measurement of the $\Omega_c(css)$ baryon. 
The invariant mass obtained by
the FOCUS Coll. is shown in Figure \ref{fig:omegacfocus}.
A summary of results on the $\Omega_c$ mass measurement is given in Table \ref{table:omegac}.

\begin{table}[htb]
\caption{Summary of $\Omega_c$ mass measurements.}
\label{table:omegac}
\newcommand{\m}{\hphantom{$-$}}
\newcommand{\cc}[1]{\multicolumn{1}{c}{#1}}
\begin{tabular}{lll}
 Coll.  &   M($\Omega_c$)[MeV]           &     decay modes                   \\ \hline 
 FOCUS  &    2697.0 $\pm$ 2.2            & $\Omega^- \pi^+$-$\Xi K \pi \pi$  \\
 Belle  &    2693.7 $\pm$ 1.3 $\pm$ 1.1  & $\Omega^- \pi^+$                  \\
 CLEO   &    2694.6 $\pm$ 2.6 $\pm$ 1.9  & ~4~decays~modes                   \\ \hline
Average &    2694.9 $\pm$ 1.3            &                                   \\ \hline
\end{tabular}\\[2pt]
\end{table}

The SELEX Coll. finds signals corresponding to a possible first observation of double-charm 
baryons  (J=1/2 ground state iso-doublet): $\Xi_{cc}^+(ccd) \rightarrow \Lambda_c^+ K^- \pi^+$ and 
$\Xi_{cc}^{++}(ccu) \rightarrow \Lambda_c^+ K^- \pi^+ \pi^+$ at a mass
of about 3520 MeV and 3460 MeV respectively. \\
This evidence is not confirmed by a similar search made by the FOCUS Coll.

\section{Heavy quark lifetimes}

The measurements of the B and D lifetimes test the decay dynamics, giving 
important information on non-perturbative QCD corrections induced by 
the spectator quark (or diquark). Decay rates are expressed using the
OPE formalism , as a sum of operators 
developed in series of order $O(\Lambda_{QCD}/m_Q)^n$. In this formalism,
no term on $1/m_Q$ is present and the spectator effects contribute at order 
$1/m_Q^3$ \footnote{Terms at order 1/$m_Q$ would appear if in this 
expansion the mass of the heavy hadron was used instead of the mass of the quark. 
The presence of this term would violate the quark-hadron duality.}.
In the B sector, non-perturbative operators are evaluated, most reliably, using 
lattice QCD calculations.

\subsection{Beauty hadron lifetimes}
Measurements of the different B hadron lifetimes have been a field of
intense experimental activity at LEP/SLD/CDF in the last ten years
and recently at B-factories (for $B^0_d$ and $B^+$ mesons only).
Results are given in Table \ref{table:life} \cite{lifeWG}.

\begin{table}[htb!]
\caption{Summary of B hadron lifetime results (as calculated by the Lifetime 
Working Group \cite{lifeWG}).}
\begin{center}
\label{table:life}
\begin{tabular}{@{}ll}
 B Hadrons                          &    ~~~~~~ Lifetime~[ps]                 \\ \hline 
~~$\tau({B^0_d})$                     & 1.540 $\pm$ 0.014 ~ (0.9~$\%$) \\ 
~~$\tau({B^+})$                       & 1.656 $\pm$ 0.014 ~ (0.8~$\%$) \\
~~$\tau({B^0_s})$                     & 1.461 $\pm$ 0.057 ~ (3.9~$\%$) \\
~~$\tau({\Lambda^0_b})$               & 1.208 $\pm$ 0.051 ~ (4.2~$\%$) \\ \hline
 \multicolumn{2}{c}{$\tau({B^0_d})/\tau({B^+})$        ~~~~~=~ 1.073 $\pm$ 0.014}                 \\
\multicolumn{2}{c}{$\tau({B^0_d})/\tau({B^0_s})$       ~~~~~~=~ 0.949 $\pm$ 0.038 }                \\
\multicolumn{2}{c}{$\tau({\Lambda^0_b})/\tau({B^0_d})$ ~~~~~~=~ 0.798 $\pm$ 0.052 }                \\ 
\multicolumn{2}{c}{$\tau({\rm{b-baryon}})/\tau({B^0_d})$       ~=~ 0.784 $\pm$ 0.034 }      \\ \hline
\end{tabular}
\end{center}
\end{table}

\begin{figure}[htb!]
\includegraphics[width=80mm]{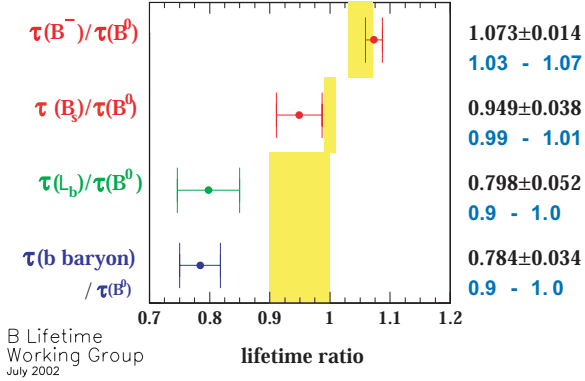}
\caption{B hadrons lifetime ratios \cite{lifeWG}, compared with theoretical predictions as given 
by the yellow bands.}
\label{fig:liferatio}
\end{figure}

Ratios of different B hadron lifetimes, given in Figure \ref{fig:liferatio},
are compared with theory predictions (yellow bands).

The attained experimental precision is remarkable and LEP results are 
still dominating the scene. The fact that charged B mesons live longer than
neutral B mesons is now established at 5$\sigma$ level and is in agreement with theory.
The $B^0_d$ and $B^0_s$ lifetimes are expected (at $\simeq$1$\%$) and found (at $\simeq$4$\%$) 
to be equal. A significant measurement in which this ratio differs 
from unity will have major consequences for the theory. 
The b-baryons lifetime is measured to be shorter than the $B^0_d$ lifetime, and the size 
of this effect seems to be more important than predicted (2-3$\sigma$).
Recent calculations of high order terms give an evaluation of 
the b-baryon lifetime 
in better agreement with the experimental result \cite{vittorio}. \\
New results are expected from B-Factories (which could decrease the relative error on 
the lifetimes of the $B^0_d$ and $B^+$ to 0.4-0.5$\%$) and mainly from Tevatron (Run II) 
which could precisely measure all B hadron lifetimes, including the $\Xi_b$, 
$\Omega_b$ and the $B_c$.

\subsection{Charm hadron lifetimes}
Differences between charm-hadron lifetimes are expected to be larger than for b-hadrons 
due to the smaller value of the charm quark mass.                                      

\begin{figure}[htb!]
\includegraphics[width=35mm,height=45mm]{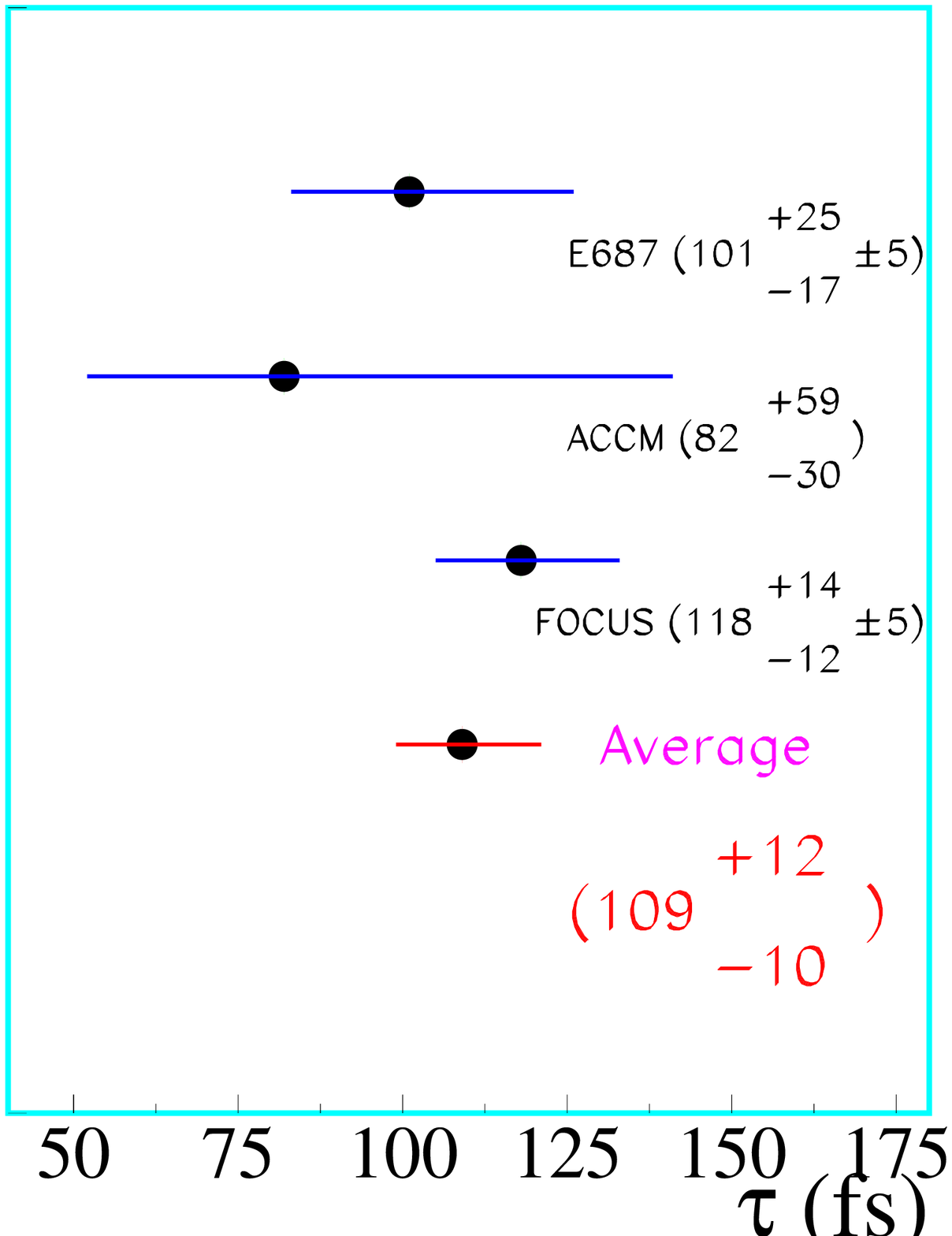}
\includegraphics[width=35mm,height=45mm]{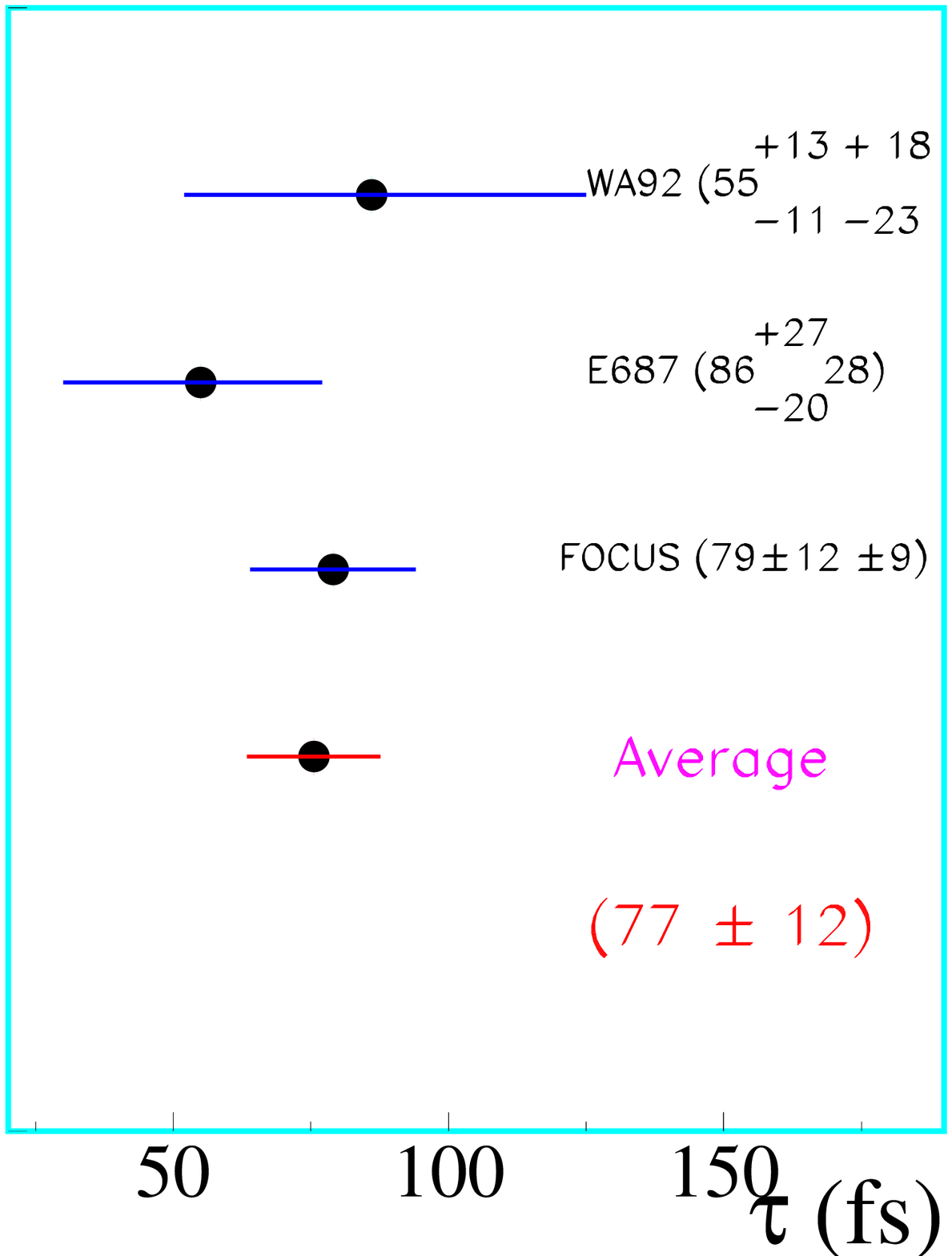}
\caption{Summary of individual $\Xi^0_c$ and $\Omega_c$ charm baryon lifetimes measurements.}
\label{fig:xi0omega0life}
\end{figure}

\begin{table}[htb!]
\caption{Summary of charm hadron lifetime results. When new results were presented at this Conference, 
averages have been made by the author.}
\label{table:dlife}
\newcommand{\m}{\hphantom{$-$}}
\newcommand{\cc}[1]{\multicolumn{1}{c}{#1}}
\begin{tabular}{@{}lll}
   Hadrons                         &    Lifetimes [fs]            &        Comments   \\ \hline  
 $\tau({D^0})$                     &  411.3 $\pm$ 1.3       & New Belle/FOCUS   \\
 $\tau({D^+})$                     & 1039.4 $\pm$ 6.3       & New Belle/FOCUS   \\  
 $\tau({D_s})$                     &  490   $\pm$ 9         &       PDG2002     \\
 $\tau({\Lambda_c})$               &  200   $\pm$ 6         &       PDG2002     \\
 $\tau({\Xi^+_c})$                 &  442   $\pm$ 26        &       PDG2002     \\
 $\tau({\Xi^0_c})$                 &  109$^{+12}_{-10}$     &  New FOCUS        \\
 $\tau({\Omega_c})$                &   77   $\pm$ 12        &  New FOCUS    \\ \hline
 \multicolumn{3}{c}{$\tau({D^+})/\tau({D^0})$    ~~=~~         2.53 $\pm$ 0.02 }                     \\
 \multicolumn{3}{c}{$\tau({D_s})/\tau({D^0})$   ~~~=~~         1.19 $\pm$ 0.02 }                \\
 \multicolumn{3}{c}{$\tau({\Lambda_c})/\tau({D^0})$ ~~~=~~     0.49 $\pm$ 0.01 }                 \\
 \multicolumn{3}{c}{$\tau({\Xi_c^+})/\tau({\Lambda_c})$ ~~~=~~ 2.21 $\pm$ 0.15 }                 \\
 \multicolumn{3}{c}{$\tau({\Omega_c})/\tau({\Xi^0_c})$  ~~~=~~ 0.71 $\pm$ 0.13 }             \\
\hline
\end{tabular}\\[2pt]
\end{table}
Important improvements have been recently made in this sector, mainly 
by the FOCUS Coll., 
producing results which are often better than previous world averages. 
New results obtained in the baryon sector are shown in Figure
\ref{fig:xi0omega0life}. A summary of charm hadron lifetime measurements is given in Table \ref{table:dlife}.\\
The charm hadron lifetime hierarchy is observed as predicted by theory.
Nevertheless, the remarkable improvement in the experimental precision 
is not yet matched by theory calculations. 

\section{Rare B decays}
 
Rare B decays were the realm of CLEO Coll., with about 9M pairs of B mesons
registered, which allowed to access B decay modes of branching
fraction of the order of 10$^{-5}$. In few areas the LEP experiments contributed too.
Since the statistics is the main issue, these studies have become 
a central topic in the B-factory program, which have now (with about 90M pairs of B mesons registered)
the possibility of accessing branching fractions of the order of 10$^{-6}$.

\subsection{Radiative B decays ( $b \rightarrow s \gamma$ )}

The radiative B decays proceed via the penguin diagrams. 
The first observation of these events was made by the CLEO Coll. in 1993 \cite{cleobsgamma}.

\begin{table*}[htb!]
\begin{center}
\caption{Summary of the results on exclusive $b \rightarrow s \gamma$ decays. Part of these results are still
preliminary and the averages have been made by the author.}
\label{table:exclbsgamma}
\newcommand{\m}{\hphantom{$-$}}
\newcommand{\cc}[1]{\multicolumn{1}{c}{#1}}
\begin{tabular}{@{}llll}
 Collaboration & $B^0 \rightarrow K^{*0} \gamma [10^{-6}]$ 
               & $B^- \rightarrow K^{*-} \gamma [10^{-6}]$     
               & $B^0 \rightarrow K_2^{*0}(1430) \gamma [10^{-6}]$ \\ \hline  
 CLEO          &  45.5 $\pm$ 7.0 $\pm$ 3.4  & 37.6 $\pm$ 8.6 $\pm$ 2.8  & 16.6 $\pm$ 5.6 $\pm$ 1.3  \\
               &  \multicolumn{2}{c}{ $A_{CP}$ = 0.08 $\pm$ 0.13 $\pm$ 0.03}        &            \\
 BaBar~~(22.7 MB)         &  42.3 $\pm$ 4.0 $\pm$ 2.2  & 38.3 $\pm$ 6.2 $\pm$ 2.2  &                           \\
          & \multicolumn{2}{c}{ $A_{CP}$ = -0.044 $\pm$ 0.076 $\pm$ 0.012}    &                 \\
 Belle~~(65.4 MB)         &  39.1 $\pm$ 2.3 $\pm$ 2.5  & 42.1 $\pm$ 3.5 $\pm$ 3.1  & 15$^{+6}_{-5}$ $\pm$ 1 (\rm{with~only ~30MB})      \\
      & \multicolumn{2}{c}{ $A_{CP}$ = -0.022 $\pm$ 0.048 $\pm$ 0.017}    &   \\ \hline
    Average             &  ~~41.4 $\pm$ 2.6      &  ~~39.8 $\pm$ 3.5    &   16.1 $\pm$ 4.2             \\ 
               &  \multicolumn{2}{c}{ $A_{CP}$ = -0.02 $\pm$ 0.04}        &            \\\hline
theory \cite{algh},\cite{bfs},\cite{bb}    &  \multicolumn{2}{c}{Br~[40-100]}       &           \\
\hline
\end{tabular}\\[2pt]
\end{center}
\end{table*}

\begin{figure}[htb!]
\includegraphics[width=80mm]{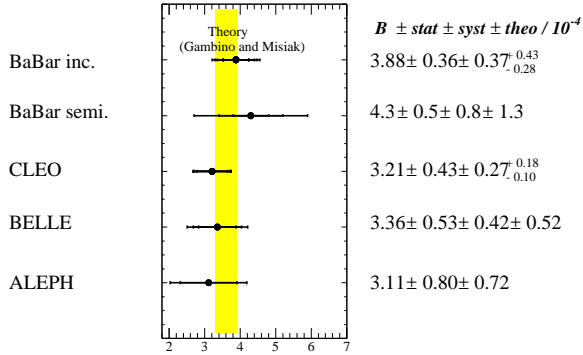} 
\caption{Summary of the individual measurements of the inclusive $b \rightarrow s \gamma$ decays 
compared with the theoretical predictions.}
\label{fig:inclbsgamma}
\end{figure}

There are three main motivations for studying radiative B decays:
\begin{itemize}
\item they are sensitive to New Physics (heavy particles in the loop);
\item the photon energy spectrum can be used to extract 
      non-perturbative QCD 
      parameters, as the b-quark mass and the Fermi motion of 
      the light quark inside the hadrons (which are important to reduce
      the error on the extraction of $V_{cb}$ and $V_{ub}$ when using inclusive b-semileptonic decay 
      samples);
\item the ratio $\frac{Br(b \rightarrow d \gamma)}{Br(b \rightarrow s \gamma)}$ 
      is proportional to the ratio of CKM matrix elements $|V_{td}/V_{ts}|^2$.
\end{itemize}

From the theoretical point of view, inclusive decays are ''cleaner`` than 
the exclusive ones, because the latter depend upon not yet well controlled form factors.
The determination of the CP asymmetry \footnote{We recall the definition
of the CP asymmetry:
$A_{CP} = \frac{Br(\bar{B} \rightarrow \bar{f}) - Br(B \rightarrow f)}
              {Br(\bar{B} \rightarrow \bar{f}) + Br(B \rightarrow f)}$},
which is expected to be small in the Standard Model ($<0.5\%$), can be a good 
place for studying non-SM CP violation.

New results from B-factories have been presented. A summary of the exclusive 
$b \rightarrow s \gamma$ decays is given in Table \ref{table:exclbsgamma}. The
measured branching fractions are compatible with the predicted ones and the 
CP asymmetry is compatible with zero within the error of about 4$\%$.\\
First results exist on $b \rightarrow d \gamma$ exclusive decays (involving $\rho^0, \rho^+, \rm{and} ~\omega$),
which combined with the results given in Table \ref{table:exclbsgamma} imply:
\begin{eqnarray}
\label{eq:bsbdgamma}
R = \frac{Br(B \rightarrow \rho \gamma)}{Br(B \rightarrow K^* \gamma)} < 0.046~~ \rm{at}~90~\%~C.L. 
\end{eqnarray}
This limit is typically a factor two (with a large error) larger than the SM 
expectations \cite{alipar},\cite{bb}\footnote{It can be also reminded that the ratio in
equation \ref{eq:bsbdgamma} is cleaner from the theoretical point of view, if only neutral 
B mesons are used for the $B \rightarrow \rho \gamma$ decay mode \cite{ref:soni1}}
(using the current determination of the Unitarity Triangle R = 0.023 $\pm$ 0.012 
is obtained \cite{ali}) and cannot yet be translated into an effective constraint on 
$|V_{td}/V_{ts}|^2 \propto (1-\bar{\rho})^2+\bar{\eta}^2$.

The BaBar Coll. presented also two new $b \rightarrow s \gamma$ inclusive 
analyses. The experimental situation, compared with the most recent theoretical calculation 
\cite{gambinomisiak} is shown in Figure \ref{fig:inclbsgamma}.
Two comments can be made: on the one hand, the agreement between experimental results and
theoretical calculations is excellent, on the other hand the experimental precision 
is approaching the theoretical uncertainties.

\subsection{Rare leptonic B decays ($B \rightarrow X_s \ell^+ \ell^-$)}
 
\begin{table*}[htb!]
\begin{center}
\caption{Summary of the results on exclusive $B \rightarrow X_s \ell^+ \ell^-$ decays. Results 
are still preliminary and averages have been made by the author. }
\label{table:kll}
\newcommand{\m}{\hphantom{$-$}}
\newcommand{\cc}[1]{\multicolumn{1}{c}{#1}}
\begin{tabular}{@{}llll}
 Collaboration  & $B \rightarrow X_s \ell \ell [10^{-7}]$ 
                & $B \rightarrow K^{*} \ell \ell [10^{-7}]$     
                & $B \rightarrow K \ell \ell [10^{-7}]$ \\ \hline  
 BaBar~~(84.4 MB) &                            & $<$ 30 at 90$\%$CL  & 7.8$^{+2.4+1.1}_{-2.0-1.8}$ \\
 Belle~~(65.4 MB) &  61 $\pm$ 14$^{+13}_{-11}$ & $<$ 14 at 90$\%$CL  & 5.8$^{+1.7}_{-1.5}$ $\pm$ 0.6  \\ \hline
    Average     &                         &                     &    6.4 $\pm$ 1.5               \\ \hline
  theory \cite{algh}  &       ~[40-50]     &     ~[10-20]    &          ~[2-5]           \\   
\end{tabular}\\[2pt]
\end{center}
\end{table*}

Motivations for the study of rare leptonic and radiative B decays are similar. 

\begin{figure}[htb!]
\includegraphics[width=70mm]{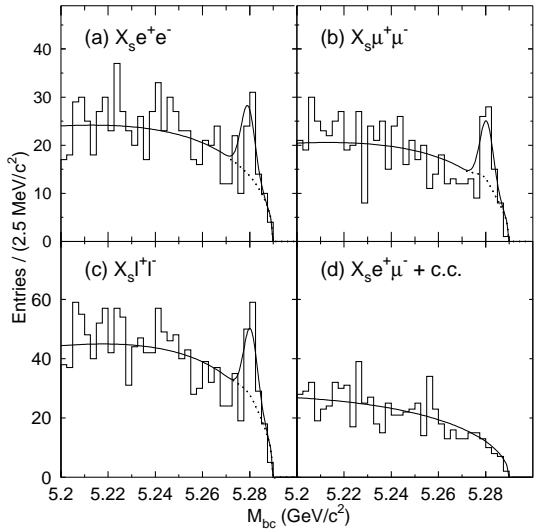} 
\caption{First observation, by Belle Coll. (preliminary result), of the inclusive decay: $b \rightarrow X_s \ell^+ \ell^-$. 
The top left(right) plots show the invariant mass spectrum for the signal in the electron(muon) 
channels, while the bottom left shows the sum of the two channels. 
The bottom right plot shows the mass spectrum for the e$\mu$ 
channel where no signal is expected.} 
\label{fig:inclxsll}
\end{figure}

A summary of exclusive leptonic B decay measurements is given in Table \ref{table:kll}.\\
The Belle Coll. has obtained the first observation of the inclusive 
$b \rightarrow X_s \ell^+ \ell^-$ decays (Figure \ref{fig:inclxsll}).

Babar Coll. presented also a limit on the $Br(B^+ \rightarrow K^+ \nu \bar{\nu}) < 9.4 \times 10^{-5}$
at 90$\%$ C.L. ( where the SM expectations is 3.8 $\times$ 10$^{-6}$)

\subsection{B hadronic decays}
Exclusive hadronic B decays are a gold mine for weak and hadronic physics.\\
One of the important goals for studying these decays is the extraction of the Unitarity Triangle angles. \\
Hadronic B decays can be schematically classified as: \\
B $\rightarrow$ Charmonium decays ;  \\
B $\rightarrow$ Open Charm decays (DX, DD,....); \\
B $\rightarrow$ Charmless B decays ($\pi \pi$ , K$\pi$....).\\
Three kinds of measurement can be performed: branching fractions, CP asymmetries ($A_{CP}$) and
time dependent CP asymmetry ( $f_{\pm}=\frac{e^{(-{\Delta t}/\tau)}}{4 \tau}
[1 \pm S_f sin(\Delta m_d \Delta t) \mp C_f cos(\Delta m_d \Delta t)]$\\
The cleanest way for extracting a weak angle is the study of the time dependence of
CP asymmetry. The ``golden channel'' is the decay mode: $B \rightarrow J/\Psi K^0$ for 
the extraction of the $\beta$ angle. The angles $\alpha$ and $\gamma$ can be, in principle, 
extracted from the study of the time dependence in charmless B decays. \\
Results from these analyses have been presented in two dedicated plenary talks from
the BaBar \cite{babartalk} and Belle \cite{belletalk} Coll.. In this paper few examples 
of experimental results are selected to show the impressive work in this field and also the richness 
of the hadronic physics informations which can be extracted.

\subsubsection{Open Charm decays}
{\underline{$B \rightarrow D \pi$ and other colour suppressed decays}}.\\ 
The study of Open Charm decays gives an important test for the B decay dynamics.
As an example, $B \rightarrow D \pi$ decay channels can be used.
\begin{table*}[htb]
\begin{center}
\caption{Summary of results on colour suppressed modes in Open Charm B decays
( common systematics from $D^0$ branching fractions can be neglected at the present level of precision). Part of these
results are still preliminary and averages have been made by the author.}
\label{table:colorsupp}
\newcommand{\m}{\hphantom{$-$}}
\newcommand{\cc}[1]{\multicolumn{1}{c}{#1}}
\begin{tabular}{@{}lllll}
 Collaboration  &  $B \rightarrow \bar{D}^0 \pi^0$    [10$^{-4}$]    & 
                   $B \rightarrow \bar{D}^0 \eta^0$   [10$^{-4}$]    &  
                   $B \rightarrow \bar{D}^0 \omega^0$ [10$^{-4}$]    & 
                   $B \rightarrow \bar{D}^0 \rho^0$   [10$^{-4}$]    \\ \hline
 CLEO           &  2.7 $^{+0.36}_{-0.32}$ $\pm$ 0.55  & - & - & -                                                 \\
 BaBar~~(~50 MB)  &  2.9 $\pm$ 0.3 $\pm$ 0.4       & 2.4 $\pm$ 0.4 $\pm$ 0.3   & 2.5 $\pm$ 0.4 $\pm$ 0.3 & -        \\
 Belle~~(~29 MB)  &  3.1 $\pm$ 0.4 $\pm$ 0.5  & 1.4 $^{+0.5}_{-0.4}$ $\pm$ 0.3 & 1.8 $\pm$ 0.5 $^{+0.4}_{-0.3}$
                                                                               & 3.0 $\pm$ 1.3 $\pm$ 0.4 (60MB)     \\
 average          & 2.93 $\pm$ 0.34           & 2.2 $\pm$ 0.4                  &  2.3 $\pm$ 0.4                   \\
\hline
\end{tabular}\\[2pt]
\end{center}
\end{table*}

\begin{table*}[htb]
\begin{center}
\caption{Summary of results on $B \rightarrow D_s \pi$ and $B \rightarrow D_s K$ decay modes. 
Results are preliminary and averages have been made by the author.}
\label{table:dspi}
\newcommand{\m}{\hphantom{$-$}}
\newcommand{\cc}[1]{\multicolumn{1}{c}{#1}}
\begin{tabular}{@{}lllll}
 Collaboration  &      $B \rightarrow D_s \pi$  [10$^{-5}$]           & $B \rightarrow D_s K$ [10$^{-5}$]      \\ \hline
 BaBar(~85 MB)  &     3.1 $\pm$ 0.9 $\pm$ 1.0                         & 3.2 $\pm$ 1.0 $\pm$ 1.0                \\
 Belle(~85 MB)  &     2.4 $^{+1.0}_{-0.8}$ $\pm$ 0.7                  & 4.6 $^{+1.2}_{-1.1}$ $\pm$ 1.3         \\
 average  & 2.7 $\pm$ 0.7 $\pm$ 0.7($D_s \rightarrow \phi \pi$) & 3.8 $\pm$ 0.9 $\pm$ 1.0($D_s \rightarrow \phi \pi$)  \\
\hline
\end{tabular}\\[2pt]
\end{center}
\end{table*}

All the $B \rightarrow D \pi$ decays rates are measured and can be described
by the color-allowed and color-suppressed diagrams (in particular the 
$B^0_d \rightarrow \bar{D}^0 \pi^0$ can proceeds only via colour-suppressed diagrams) . 
The amplitudes can be expressed in terms of isospin amplitudes (I=1/2 and I=3/2) and of their relative
strong phase shift ($\delta_I = \delta_{I=3/2}-\delta_{I=1/2}$).
Using results from Belle and CLEO collaborations it results: 
\begin{eqnarray}
\label{eq:cosdelta}
                 cos \delta_I &=& 0.866^{+0.042}_{-0.036} 
\end{eqnarray}
which is at 3.2$\sigma$ different from unity and indicates sizable final-states re-scattering 
effects in $D \pi$ decays.
Other colour-suppressed modes are now measured and are summarised in Table \ref{table:colorsupp}.
The rates are in general twice larger as those expected in the naive factorization approach.

\noindent
{\underline{$B \rightarrow D_s \pi$ and $B \rightarrow D_s K$}}.
\begin{figure}[htb!]
\includegraphics[width=75mm]{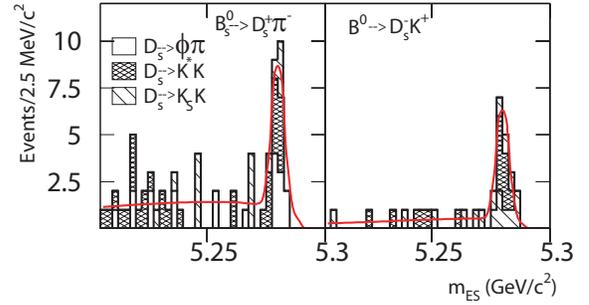} 
\caption{The mass spectrum for the first preliminary observation of the $B \rightarrow D_s \pi$ and $B \rightarrow D_s K$
decay modes from the BaBar Coll..}
\label{fig:dspi}
\end{figure}
The $B \rightarrow D_s \pi$ decay mode is expected to proceed via the $b \rightarrow u$ transition, with no
penguin contribution and can provide, in principle, a way to determine $|V_{ub}|$ \cite{dspi}. It seems,
nevertheless, difficult to extract $|V_{ub}|$ with a precision better than 30$\%$. This mode can be
used to determine the ratio $R_{\lambda} = \frac{A({B}^0 \rightarrow D^{(*)+} \pi^-)}
{A({B}^0 \rightarrow D^{(*)-} \pi^+)}$ which is important for a possible extraction of sin2$(\beta +\gamma)$ 
from the study of the time evolution of the $B^0 \rightarrow D^{(*)-} \pi^+$ decays. Unfortunately the
mode ${B}^0 \rightarrow D^{*+} \pi^-$ is difficult to access experimentally (background
$\bar{B}^0 \rightarrow D^{*+} \pi^-$). The $B \rightarrow D_s \pi$ can be then used relating the mode
${B}^0 \rightarrow D^{*+} \pi^-$ to  ${B}^0 \rightarrow D_s^{*+} \pi^-$ using SU(3) symmetry.\\
The decay $B \rightarrow D_s K$ can occur via W-exchange or final state re-scattering and cannot be described
by a spectator graph. Wide ranges of prediction of its branching ratio exist (from 3$\times$ 10$^{-6}$ 
to 10$^{-4}$ \cite{dsk}).
First results on $B \rightarrow D_s \pi$ and $B \rightarrow D_s K$ are shown in Figure \ref{fig:dspi} and 
summarised in Table \ref{table:dspi}. The uncertainty on $Br(D_s \rightarrow \phi \pi)$ is already limiting 
the precision on the $Br(B \rightarrow D_s \pi)$.
 
\subsubsection{Charmless B decays}
\begin{table*}[htb!]
\caption{Summary of results on the charmless B decays. Many of these results are preliminary and obtained using
a registered statistics which depend upon the experiments and the analysed
decay modes.}
\label{table:charmless}
\newcommand{\m}{\hphantom{$-$}}
\newcommand{\cc}[1]{\multicolumn{1}{c}{#1}}
\scriptsize{
\begin{tabular}{@{}llllll}
     Decay Mode   &  Br [10$^{-6}$]  &  Br [10$^{-6}$]  &  Br [10$^{-6}$]  &    A$_{CP}$   & A$_{CP}$  \\
                  &    (Belle)       &      (BaBar)     &       (CLEO)     &    (Belle )  & (BaBar) \\ \hline
$B^0 \rightarrow K^+ \pi^-$ & 22.5 $\pm$ 1.9 $\pm$ 1.8  &  17.9 $\pm$ 0.9 $\pm$ 0.6  & 17.2$^{+2.5}_{-2.4}$ $\pm$ 1.2& 
                                                 -0.06 $\pm$ 0.09 $^{+0.01}_{-0.02}$ & -0.102$\pm$0.050$\pm$ 0.016   \\
$B^0 \rightarrow K^+ \pi^0$ & 13.0$^{+2.5}_{-2.4}$ $\pm$ 1.3  & 12.8 $\pm$ 1.2 $\pm$ 1.0 & 11.6$^{+3.0+1.4}_{-2.7-1.3}$ & 
                                                   -0.02 $\pm$ 0.19 $\pm$ 0.02        & -0.09$\pm$0.09$\pm$ 0.01        \\
$B^0 \rightarrow \pi^+ \pi^0$ & 7.4$^{+2.3}_{-2.2}$ $\pm$ 0.9  &  5.5 $\pm$ 1.0 $\pm$ 0.6  & 5.4$^{+2.1}_{-2.0}$ 
                                                                                                             $\pm$ 1.5  & 
                                                    0.30 $\pm$ 0.30 $^{+0.06}_{-0.04}$  & -0.03$\pm$0.18$\pm$ 0.02     \\
$B^0 \rightarrow K^0 \pi^0$ & 8.0$^{+3.3}_{-3.1}$ $\pm$ 1.6  & 10.4 $\pm$ 1.5 $\pm$ 0.8 & 14.6$^{+5.9+2.4}_{-5.1-3.3}$ & 
                                                                                         & 0.03$\pm$0.36$\pm$ 0.09     \\
$B^0 \rightarrow \pi^+ \pi^-$ & 5.4 $\pm$ 1.2 $\pm$ 0.5  &  4.6 $\pm$ 0.6 $\pm$ 0.2  & 4.3$^{+1.6}_{-1.4}$ $\pm$ 0.5  & 
  S=-1.21$^{+0.38+0.16}_{-0.27-0.13}$  & S=0.02 $\pm$ 0.34 $\pm$ 0.05  \\
& & & & C=0.94$^{+0.25}_{-0.31}$$\pm$ 0.09 & C=-0.30 $\pm$ 0.25 $\pm$ 0.04 \\

$B^0 \rightarrow K^0 \bar{K^0}$ & $<$4.1 &   $<$7.3  & $<$ 13 & &  \\
$B^0 \rightarrow \pi^0 \pi^0$   & $<$6.4 &   $<$3.6  & $<$ 5.2 & &  \\
$B^+ \rightarrow K^0 \pi^+$ & 19.4$^{+3.1}_{-3.0}$ $\pm$ 1.6 & 17.5 $\pm$ 1.8 $\pm$ 1.3 & 18.2$^{+4.6}_{-4.0}$  $\pm$ 1.6                                                           &  0.46 $\pm$ 0.15$ \pm$ 0.02 & -0.17 $\pm$ 0.10 $\pm$ 0.02 \\
$B^+ \rightarrow K^+ {K^0}$   & $<$2.0 &   $<$1.3  & $<$ 5.1 & &  \\
$B^+ \rightarrow K^+ {K^-}$   & $<$0.9 &   $<$0.6  & $<$1.9  & &  \\
\hline
$B^+ \rightarrow \rho^0 \pi^+$ & 8.0$^{+2.2}_{-2.0}$ $\pm$ 0.7  
& 24$\pm$8$\pm$3  
& 10.4 $^{+3.3}_{-3.4}$ $\pm$ 2.1   &
                                                                         &      \\
$B^0 \rightarrow \rho^{\pm} \pi^{\mp}$ & 20.8$^{+6.0+2.8}_{-6.3-3.1}$  &  28.9 $\pm$ 5.4 $\pm$ 4.3 
                                                                               &  27.6$^{+8.4}_{-7.4}$ $\pm$ 4.2  & 
                                                                              & -0.22$\pm$0.08 $\pm$ 0.07\\
$B^0 \rightarrow \rho^0 \pi^0$   & $<$5.3 &   $<$10.6  & $<$ 5.5 & &  \\
$B^0 \rightarrow a_1^{\pm} \pi^{\mp}$   &  &   6.2$^{+3.0}_{-2.5}$ $\pm$ 1.1  &  &  &   \\
$B^+ \rightarrow \rho K$               & $<$12 &   10.7 $\pm$ 1.0 $^{+0.9}_{-1.6}$  &  &  &0.19 $\pm$ 0.14 $\pm$ 0.11 \\
$B^+ \rightarrow \eta^{'} K^+$ & 77.9$^{+6.2+9.3}_{-5.9-8.7}$  &  67 $\pm$ 5 $\pm$ 5 
                                                                                   &  80$^{+10}_{-9}$ $\pm$ 7  & 
                                                     -0.015 $\pm$ 0.070 $\pm$ 0.009  & -0.11 $\pm$ 0.11 $\pm$ 0.02 \\
$B^0 \rightarrow \eta^{'} K^0$ & 68.0$^{+10.4+8.8}_{-9.6-8.2}$  &  46 $\pm$ 6 $\pm$ 4 
                                 &  89$^{+18}_{-16}$ $\pm$ 9  & 0.26$\pm$ 0.22 $\pm$0.03 &  \\
$B^+ \rightarrow \eta^{'} \pi^+$ &  & 5.4$^{+3.5}_{-2.6}$ $\pm$0.8  & $<$12  &  &  \\
$B \rightarrow \eta^{'} K^{*0}(K^{*+})$   & $<$20(90) &   $<$13  & $<$ 24(35) & &  \\
$B \rightarrow \eta^{'} \rho^0$   & $<$14 &    & $<$ 12 & &  \\
$B^+ \rightarrow \eta K^{*+}$ & 26.5$^{+7.8}_{-7.0}$ $\pm$ 1.7  &  22.1$^{+11.1}_{-9.2}$ $\pm$ 3.3 
                                                                                   &  26.4$^{+9.6}_{-8.2}$ $\pm$ 3.3  & 
                                                                      &               \\
$B \rightarrow \eta K^{*0}$ & 16.5$^{+4.6}_{-4.2}$ $\pm$ 1.2  &  19.8$^{+6.5}_{-5.6}$ $\pm$ 1.7 
                                                                                   &  13.8$^{+5.4}_{-4.6}$ $\pm$ 1.6  & 
                                                              &              \\
$B \rightarrow \eta K^+$   & 5.3$^{+1.8}_{-1.5}$$\pm$0.6  &  & $<$6.9  & &  \\
$B \rightarrow \eta \pi^+$ & 5.4$^{+2.0}_{-1.7}$$\pm$0.6  &  & $<$5.7     & &  \\
$B \rightarrow \eta \rho^+(\rho^0)$   & $<$2.7($<$6.2)  &    & $<$15($<$10) &  &  \\
$B^{\pm} \rightarrow \omega \pi^{\pm}$  & 4.2$^{+2.0}_{-1.8}$ $\pm$ 0.5  &  6.6$^{+2.1}_{-1.8}$ $\pm$ 0.7 
                                                                                   &  11.3$^{+3.3}_{-2.9}$ $\pm$ 1.4 & 
                                                               &  -0.01$^{+0.29}_{-0.31}$ $\pm$0.03 \\
$B^{\pm} \rightarrow \omega K^{\pm}$  & 9.2$^{+2.6}_{-2.3}$ $\pm$ 1.0 & $<$4 & $<$7.9 
& -0.21 $\pm$ 0.28 $\pm$ 0.03 & \\
$B^{0} \rightarrow \omega K^0$  &  & 5.9$^{+1.7}_{-1.5}$ $\pm$ 0.9  &  $<$21 &   &  \\
$B^{0} \rightarrow \omega \pi^0$  & & $<$3.0  &  $<$5.5 &  &   \\
$B^{\pm} \rightarrow \phi K^{\pm}$  & 7.2$^{+1.5}_{-1.4}$$\pm$0.9$\pm$0.4 &  9.2 $\pm$ 1.0 $\pm$ 0.8 
                                                                                   &  5.5$^{+2.1}_{-1.8}$ $\pm$ 0.6 & 
                                                               &  -0.05$\pm$0.20$\pm$0.03 \\ \hline
$B^{\pm} \rightarrow \phi K^{*\pm}$  &  &  9.7$^{+4.2}_{-3.4}$ $\pm$ 1.7 
                                                                                   &  $<$22.5 & 
                                                               &  -0.43$^{+0.36}_{-0.30}$ $\pm$ 0.06 \\
$B^0 \rightarrow \phi K^{*0}$  &   &  8.6$^{+2.8}_{-2.4}$ $\pm$ 1.1 
                                                                   & 11.5 $^{+4.5+1.8}_{-3.7-1.7}$  & 
                                                               &  0.00$\pm$0.27$\pm$0.03 \\
$B^0 \rightarrow \phi K_S^{0}$  & 10.0 $^{+1.9+0.9}_{-1.7-1.3}$  &  8.7$^{+1.7}_{-1.5}$ $\pm$ 0.9 
   &$<$12.3                   & -0.56$\pm$0.41$\pm$0.12 & \\
$B^{\pm} \rightarrow \phi \pi^{\pm}$  &  & $<$0.56 & & & \\
$B^+ \rightarrow \rho^+ \rho^0$  & 38.5$\pm$10.9$^{+5.9+2.5}_{-5.4-7.5}$  &  & & & \\
\hline
\end{tabular}}
\end{table*}

\begin{table*}[htb]
\caption{Summary of preliminary results on $\bar{\Lambda}$ and $\lambda_1$. 
The second and third errors correspond, respectively, to the systematic and theoretical 
uncertainties ($\alpha_s$ and $1/m_b^3$).} 
\label{table:moments}
\newcommand{\m}{\hphantom{$-$}}
\newcommand{\cc}[1]{\multicolumn{1}{c}{#1}}
\begin{tabular}{@{}lll}
 Collaboration  &     $\bar{\Lambda}$                            &               $\lambda_1$         \\ \hline 
 BaBar          &     0.35 $\pm$ 0.07                            &   -0.17 $\pm$ 0.06 $\pm$ 0.07                    \\
 CLEO           &     0.39 $\pm$ 0.03 $\pm$ 0.06 $\pm$ 0.12 GeV  &   -0.25 $\pm$ 0.02 $\pm$ 0.05 $\pm$ 0.14 GeV$^2$  \\
 DELPHI         &     0.44 $\pm$ 0.04 $\pm$ 0.05 $\pm$ 0.07 GeV  &   -0.23 $\pm$ 0.04 $\pm$ 0.05 $\pm$ 0.08 GeV$^2$  \\
\hline
\end{tabular}\\[2pt]
\end{table*}

One of the interest of measuring charmless B decays is the determination of the
unitarity triangle angles $\alpha$ and $\gamma$. In general a given decay mode is described by
various tree (T) and penguin (P) diagrams which depend upon weak and strong phases.
A lot of theoretical investigations have been recently made (\cite{bbns}-\cite{othercharmless}).
In particular important progress has been made in the last years with the calculation of 
amplitudes in the heavy quark limit \cite{bbns}, but there is still some controversy on the corrections to 
it \cite{roma}. Analyses have been made, essentially on $B \rightarrow PP$ (P=pseudosclar) with contradictory 
results.\\
From the experimental point of view an impressive effort has been made
 to measure 
as many branching fractions and CP asymmetries as possible. Results are given in Table 
\ref{table:charmless}. Those on branching fractions are in fairly good agreement among different
experiments. CP asymmetries are all compatible with zero. The only ``anomaly'' is the CP asymmetry in 
the $\pi^+ \pi^-$ channel reported by the Belle Coll..

\section{Determination of $|V_{cb}|$}
\begin{figure}[htb!]
\includegraphics[width=75mm]{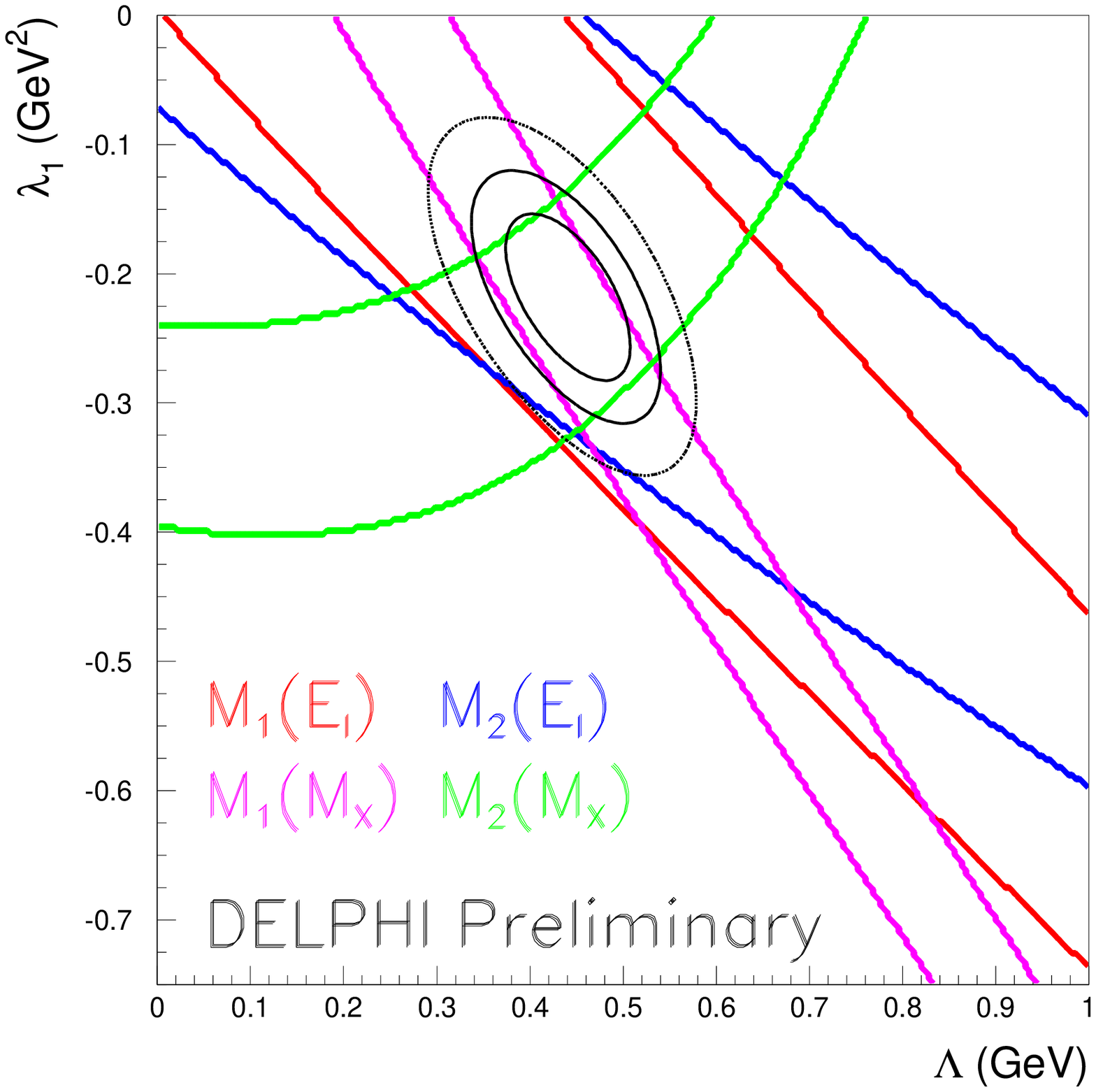} \\ 
\includegraphics[width=75mm]{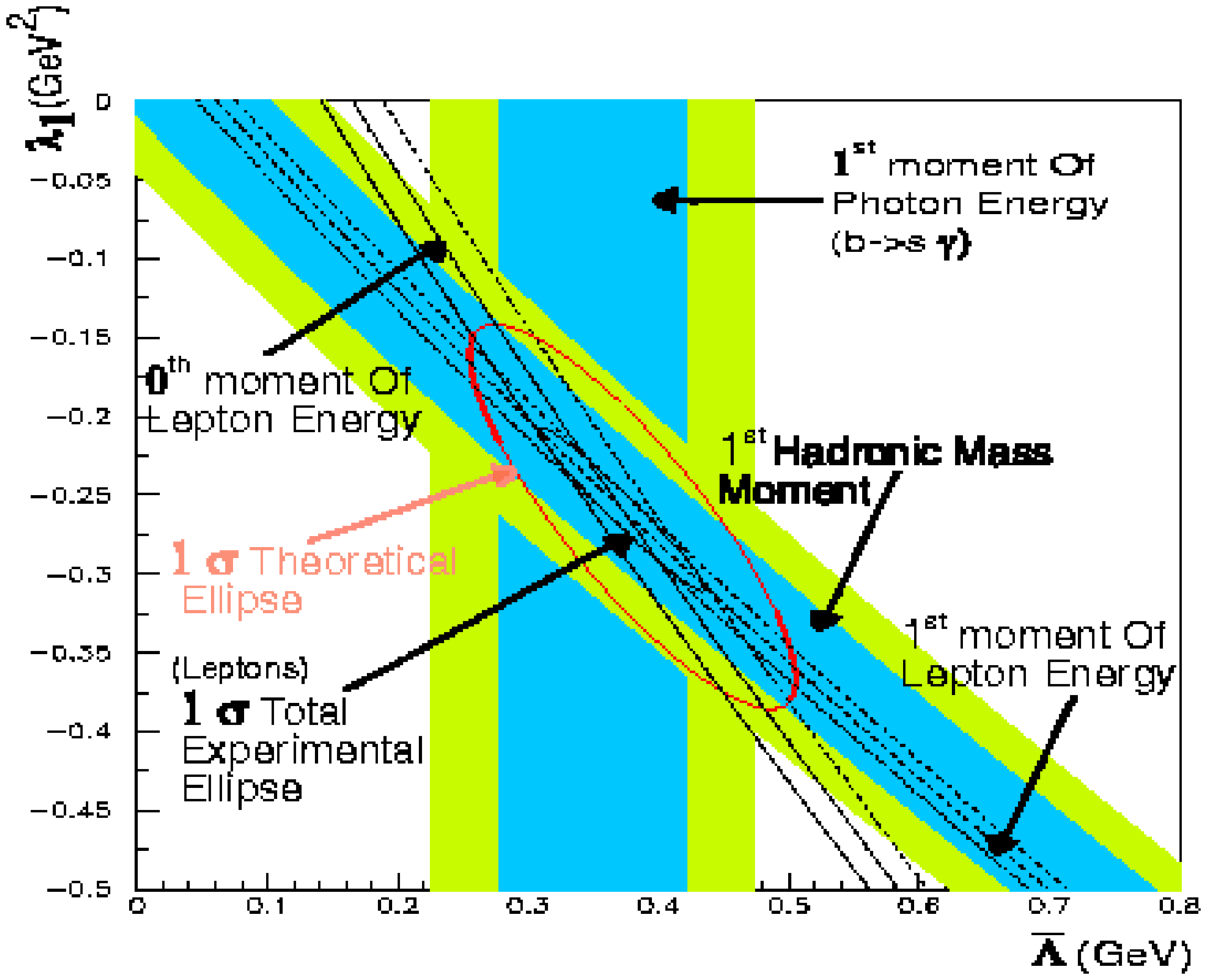} 
\caption{Constraints in the $\bar{\Lambda}-\lambda_1$ plane obtained: 
by the DELPHI Coll.
using the measured values of the first two moments of the hadronic mass and lepton energy spectra (top).
The bands represent the 1$\sigma$ regions selected by each moment and the ellipses show the 
39$\%$, 68$\%$ and 90$\%$ probability regions of the global fit ; by CLEO Coll. using the first 
moment of hadronic mass, lepton energy and $\gamma$ energy distributions (bottom). }
\label{fig:moments}
\end{figure}

The $|V_{cb}|$ element of the CKM matrix can be accessed by studying the decay 
rates of inclusive and exclusive semileptonic $b$-decays.
\subsection{Inclusive analysis.}
The first method to extract $|V_{cb}|$ makes use of the inclusive
semileptonic decays of B-hadrons and of the theoretical calculations
done in the framework of the OPE.
The inclusive semileptonic width $\Gamma_{s.l.}$ is expressed as:
\begin{eqnarray} 
 \Gamma_{s.l.} =  \frac{BR(b \rightarrow c l \nu)}{\tau_b}  = \gamma_{theory} |V_{cb}|^2 ; & \nonumber \\
  \gamma_{theory} = f(\alpha_s,m_b,\mu_{\pi}^2,1/m_b^3...).
\label{eq:vcbtheo}
\end{eqnarray} 
From the experimental point of view the semileptonic width has been measured by the LEP/SLD and 
$\Upsilon(4S)$ experiments with a relative precision of about 2$\%$:
\begin{eqnarray}
 \Gamma_{sl} = (0.431 \pm 0.008 \pm 0.007) 10^{-10} ~MeV  \quad  \small{\Upsilon(4S)}  \nonumber \\
 \Gamma_{sl} = (0.439 \pm 0.010 \pm 0.007) 10^{-10} ~MeV  \quad \small{\rm{Z^0}} \nonumber 
\label{eq:gammsl1} 
\end{eqnarray}
The average is :
\begin{equation}
 \Gamma_{sl} = (0.434 \times (1 \pm 0.018)) 10^{-10} ~MeV. 
\label{eq:gammsl2} 
\end{equation}

The precision on the determination of $|V_{cb}|$ is mainly limited by theoretical 
uncertainties on the parameters entering in the expression of $\gamma_{theory}$ in equation \ref{eq:vcbtheo}.

\subsection{Moments analyses}
Moments of the hadronic mass spectrum, of the lepton energy spectrum and of 
the photon energy in the $b \rightarrow s \gamma$ decay are sensitive to the non
perturbative QCD parameters contained in the factor $\gamma_{theory}$ of 
equation \ref{eq:vcbtheo} and in particular to the mass of the $b$ and $c$ quarks 
and to the Fermi motion of the light quark inside the hadron, $\mu_{\pi}^2$ \footnote{In another formalism,
based on pole quark masses, the $\bar{\Lambda}$ and $\lambda_1$ parameters are used, which can
be related to the difference between hadron and quark masses and to $\mu_{\pi}^2$, respectively.}.\\
Preliminary results, obtained by BaBar, CLEO and DELPHI Coll., are summarised in Table \ref{table:moments}
and in Figure \ref{fig:moments}.
By using the experimental results on $\bar{\Lambda}$ and $\lambda_1$ it 
gives:
\begin{equation}
 |V_{cb}| = (40.7 \pm 0.6 \pm 0.8(\rm{theo.})) 10^{-3} \rm{(inclusive)}
\label{eq:vcbinclres}
\end{equation}
This result is an important improvement on the determination of the $|V_{cb}|$ element. Part of the
theoretical errors (from $m_b$ and $\mu_{\pi}^2$) is now absorbed in the experimental error and the theoretical error is 
reduced by a factor two. The remaining theoretical error could be further reduced if the 
parameters controlling the $1/m_b^3$ corrections are extracted directly from experimental data.

\subsection{$B \rightarrow D^* \ell \nu$ analysis.}
An alternative method to determine $|V_{cb}|$ is based on exclusive $\overline{ B^0_d} \rightarrow D^{*+} \ell^- 
\overline{\nu_l}$ decays. Using HQET an expression for the differential decay rate can be derived
\begin{equation}
\frac{d\Gamma}{dw} = \frac{G_F^2}{48 \pi^2} |V_{cb}^2| |F(w)|^2 G(w) ~;~ w = v_B.v_D 
\label{eq:exclvcb}
\end{equation}
\begin{figure}[htb]
\includegraphics[width=75mm]{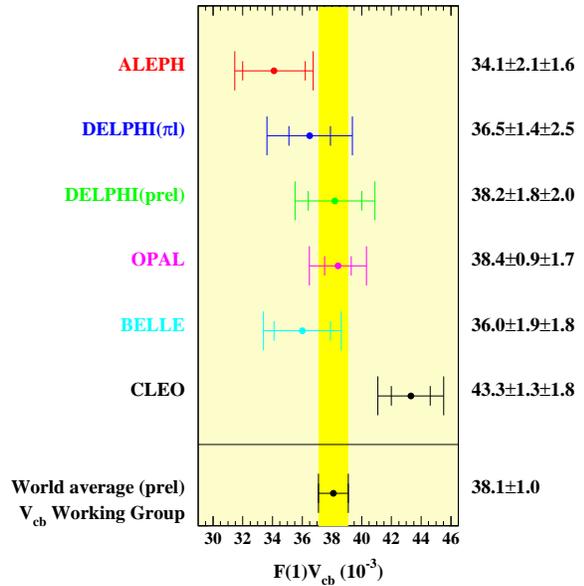}
\caption{Summary of the measurements of $F(1) \times |V_{cb}|$ \cite{vcbWG}.}
\label{fig:vcb_ave}
\end{figure}
$w$ is the relative velocity between the B ($v_B$) and the D systems ($v_D$). G($w$) is a kinematical
factor and F($w$) is the form factor describing the transition. At zero recoil ($w$=1)
F(1) goes to unity. The strategy is then to measure $d\Gamma/dw$, to extrapolate 
at zero recoil and to determine $F(1) \times |V_{cb}|$.\\
The experimental results are summarised in Figure \ref{fig:vcb_ave}.
Using F(1) = 0.91 $\pm$ 0.04 \cite{latticeF1}, it gives \cite{vcbWG}:
\begin{equation}
 |V_{cb}| = (41.9 \pm 1.1 \pm 1.9(F(1)) 10^{-3}  \rm{(exclusive)}
\label{eq:vcbexclres}
\end{equation}
Combining the two determinations of $|V_{cb}|$ (a possible correlation 
between the two determinations has been neglected) it gives:
\begin{eqnarray}
 |V_{cb}| = (40.9 \pm 0.8) 10^{-3}   \tiny{\rm{(exclusive+inclusive)}}
\label{eq:vcbave}
\end{eqnarray}

\section{Determination of $|V_{ub}|$}

This measurement is rather difficult because one has to suppress 
the large background from the more abundant semileptonic $b$ to $c$ quark transitions. \\
Several new determinations of the CKM element $|V_{ub}|$ have been presented at this Conference \cite{marco}.

\begin{figure}[htb]
\begin{center}
\includegraphics[width=75mm]{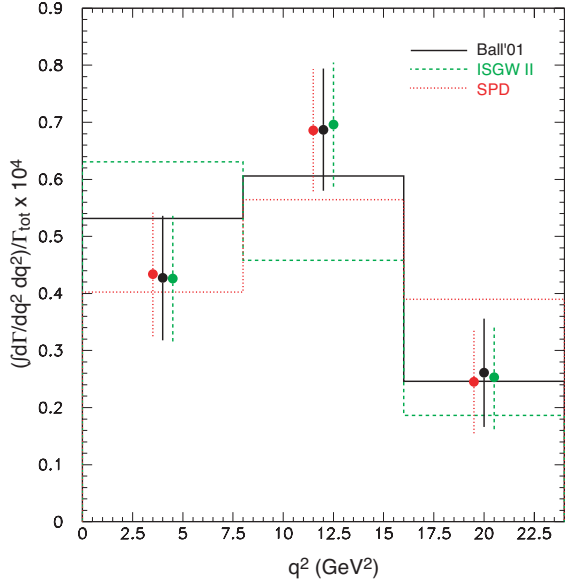}
\caption{Differential branching fraction for $B^0 \to \pi^- \ell^+ \nu$  as a function of $q^2$ by the CLEO Coll., 
compared with the predicted values (histograms) for three models used to extract $|V_{ub}|$.}
\end{center}
\label{fig:cleovub}
\end{figure}

\subsection{Determination of $|V_{ub}|$ using exclusive analyses.}

\begin{figure}[htb]
\includegraphics[width=80mm]{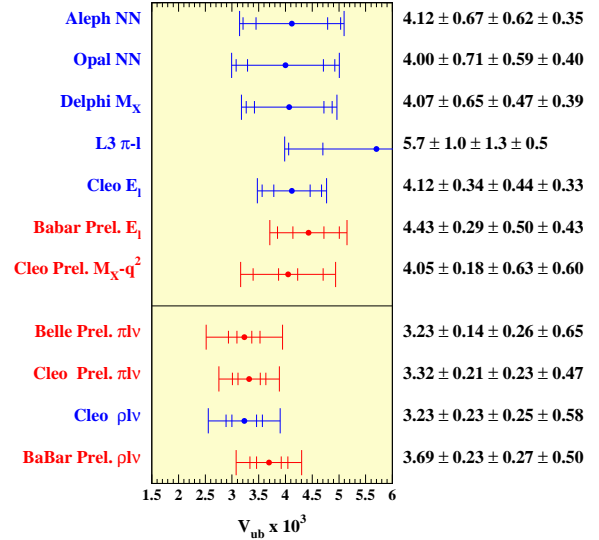}
\caption{Summary of $|V_{ub}|$ measurements \cite{marco}.}
\label{fig:vub_ave}
\end{figure}

The first method to determine $|V_{ub}|$ consists in the reconstruction 
of the charmless semileptonic B decays: $B \rightarrow \pi (\rho) \ell \nu$.\\
From the theoretical point of view, the main problem is the determination of the 
form factors. Light-Cone Sum Rules can provide an evaluation at the 15-20$\%$ level.
Lattice QCD calculations give a similar precision but these uncertainties are expected to be reduced 
in the near future. The main limitation in these calculations is that,
at present, they can be used only in the high $q^2$ region.\\
An interesting analysis has been presented by the CLEO Coll. using the $B^0 \to \pi^- \ell^+ \nu$ 
decay mode, extracting the signal rates in three independent regions of $q^2$. 
In this way it is possible to discriminate between models and the fit in Figure \ref{fig:cleovub}
shows that the ISGW~II model is compatible with data at only  1$\%$ level of probability.\\

\subsection{Determination of $|V_{ub}|$ using inclusive analyses.}
As for $|V_{cb}|$, the extraction of $|V_{ub}|$ from inclusive semileptonic 
decays is based on HQET implemented through OPE. \\
By using kinematical and topological variables, it is possible to select samples
enriched in $b \rightarrow u$ transitions. There are, schematically, three 
main regions in the semileptonic decay phase space:
\begin{itemize}
\item  the lepton energy end-point region: $E_{\ell}>\frac{M^2_B-M^2_D}{2M_B}$ (which was at the origin 
       for the first evidence of $b \rightarrow u$ transitions)
\item  the low hadronic mass region: $M_{X} < M_{D}$  (pioneered by the DELPHI Coll.)
\item  the high $q^2$ region: $M^2_{\ell \nu}=q^2>(M_B-M_D)^2$. 
\end{itemize}

The CLEO Coll. has presented an interesting attempt of a combined $M_X-q^2$ analysis
to reduce theoretical uncertainties. \\
A summary of the different determinations of $|V_{ub}|$ is given in Figure \ref{fig:vub_ave}.
It is probably too early to make an overall average using all results.

\section{Measurement of $B^0-\bar{B^0}$ oscillations}

\begin{figure}[htb]
\includegraphics[width=75mm]{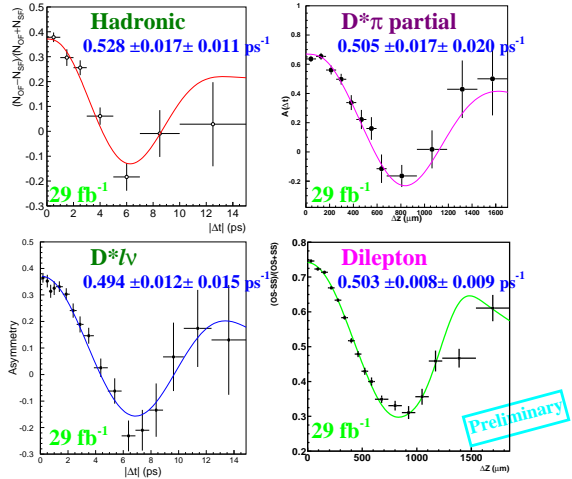}
\caption{The plots show the ${B}^0_d - \overline{{B}}^0_d$ oscillations (Belle Coll.). 
The points with error bars are the data. The result of the fit gives the value 
for $\Delta m_d$.}
\label{fig:dmdbfactory}
\end{figure}

The probability that a ${B}^0$ meson oscillates into a 
$\overline{{B}}^0$ or stays as a ${B}^0$ is given by:
\begin{equation}
P_{{B}^0_q \rightarrow {B}^0_q(\overline{{B}}^0_q)} =
\frac{1}{2}e^{-t/\tau_q} (1 \pm cos \Delta {m}_q t)
\label{eq:osci}
\end{equation}
where $\tau_q$ is the lifetime of the ${B}^0_q$ meson, $\Delta {m}_q = {m}_{B^0_1}
- m_{{B}^0_2}$ is the mass difference between the two mass eigenstates
\footnote{$\Delta {m}_q$ is usually given in ps$^{-1}$. 1 ps$^{-1}$ corresponds to 
6.58 10$^{-4}$eV.}. To derive this formula the effects of a CP violation and of the 
lifetime difference between the two states have been neglected. \\
 In the Standard Model,  $B^0-\bar{B}^0$ oscillations 
occur through a second-order process - a box diagram - with a loop of W and up-type quarks. 
 The box diagram with the exchange of a $top$ quark gives the dominant contribution.
\begin{figure}[htb]
\includegraphics[width=75mm]{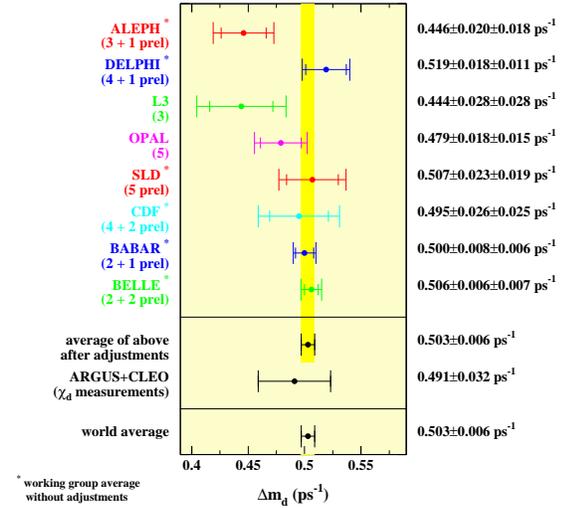}
\caption{Summary of the measurements of $\Delta m_d$ \cite{osciWG}.}
\label{fig:dmd_ave}
\end{figure}

The oscillation frequency is predicted to be:


\begin{eqnarray}
\Delta {m}_d & ~\propto~V_{cb}^2 \lambda^2 [(1-\bar{\rho})^2+\bar{\eta}^2]f^2_{{B}_d} B_{{B}_d} &\nonumber \\
\Delta {m}_s & ~\propto~V_{cb}^2 f^2_{{B}_s} B_{{B}_s}  &\nonumber \\
\frac{\Delta m_d}{\Delta m_s} & ~\propto~ 1/\xi^2 \lambda^2 [(1 - \bar{\rho})^2 + \bar{\eta}^2] &
\label{eq:dmddmsxi}
\end{eqnarray}

where  $\xi=\frac{ f_{B_s}\sqrt{B_{B_s}}}{ f_{B_d}\sqrt{B_{B_d}}}$.

Thus, the measurement of $\Delta m_d$ and $\Delta m_s$ gives access to 
the CKM matrix elements $|V_{td}|$
and $|V_{ts}|$ respectively.
The difference in the $\lambda$ dependence of these expressions ($\lambda \sim 0.22$) implies that 
$\Delta {m}_s \sim 20 ~\Delta {m}_d$. It is then clear that a very good proper time resolution is needed 
to measure the $\Delta {m}_s$ parameter.
On the other hand the measurement of the ratio $\Delta m_d/\Delta m_s$ gives the same 
constraint as $\Delta m_d$ but this ratio is expected to have smaller theoretical uncertainties 
since the ratio $\xi$ is better known than the absolute value of $f_B \sqrt B_B$.

\subsection{Measurement of the $B^0_d-\bar{B^0_d}$ oscillation frequency: $\Delta m_d$}
The measurement of $\Delta m_d$ has been the subject of intense experimental activity
in the last ten years. Results are available from the
combination of more than 35 analyses, using different event samples, performed by
the LEP/SLD/CDF experiments. The combined measurement of $\Delta m_d$ have 
a relative precision of $\sim$ 2.5$\%$.
The new and precise measurements performed at B-Factories are now included,
improving the precision by a factor of two.\\
A typical proper time distribution is shown in Figure \ref{fig:dmdbfactory}. The 
oscillating behaviour 
is clearly visible. Figure \ref{fig:dmd_ave} gives the results for $\Delta m_d$, obtained by each experiment and
the overall average \cite{osciWG}:
\begin{eqnarray}
\Delta {m}_d &=& (0.503 \pm 0.006)~{ps}^{-1}.
\label{eq:dmdresults}
\end{eqnarray}

Improvements can still be expected from B-factories which should reach a few per mille precision on $\Delta m_d$.\\

\noindent
\subsection{Search for $B^0_s-\bar{B}^0_s$ oscillations.}

\begin{figure}[htb]
\includegraphics[width=75mm]{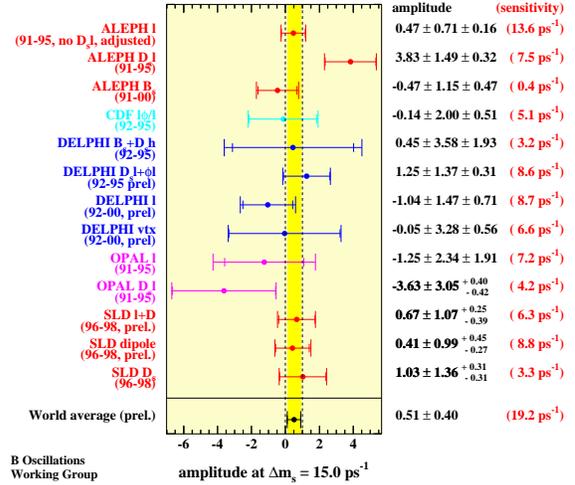}
\caption{$B_s^0$ oscillation results. Values of the fitted amplitude at $\Delta {m}_s = 15~\rm{ps}^{-1}$ and of  the sensitivity obtained by each experiment \cite{osciWG}.}
\label{fig:dms_ave}
\end{figure}

\begin{figure}[htb]
\includegraphics[width=75mm]{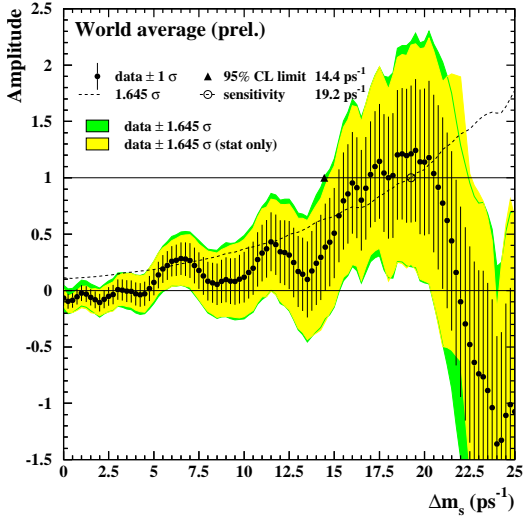}
\caption{The plot \cite{osciWG} shows combined $\Delta {m}_s$ results from 
LEP/SLD/CDF analyses shown as an amplitude versus $\Delta {m}_s$ plot. 
The points with error bars are the data; the lines show the 95\% C.L. curves (in dark the
systematics have been included). The dotted curve shows the sensitivity.}
\label{fig:dms_ampli}
\end{figure}

Since the $B_s^0$ meson is expected to oscillate more than 20 times faster than the $B_d^0$
and as $B_s$ mesons are less abundantly produced, the search for  $B^0_s-\bar{B}^0_s$ oscillations 
is more difficult. The observation of fast oscillations requires the highest resolution on the proper 
time and on the $B_s$ decay length. No signal for $B^0_s-\bar{B}^0_s$ oscillations has been observed 
sofar. \\
The method used to measure or to put a limit on $\Delta {m}_s$ consists
in modifying equation \ref{eq:osci} in the following way \cite{ampli}: $1 \pm cos \Delta {m}_s t 
\rightarrow 1 \pm A cos \Delta {m}_s t$. A and $\sigma_A$ are measured at fixed values  
of $\Delta {m}_s$ instead of $\Delta {m}_s$ itself. 
In case of a clear oscillation signal, at a given frequency, the amplitude should be 
compatible with A = 1 at this frequency.
With this method it is easy to set a limit. The values of $\Delta
{m}_s$ excluded at 95\% C.L. are those satisfying the condition A($\Delta{m}_s$) 
+ 1.645 $\sigma_A (\Delta {m}_s) < 1$. 
Furthermore the sensitivity of the experiment can be defined as the value of
$\Delta {m}_s$ corresponding to 1.645 $\sigma_A (\Delta {m}_s) = 1$ (taking
A($\Delta {m}_s) = 0$), namely supposing that the ``true'' value of
$\Delta {m}_s$ is well above the measurable value. \\
During the last years impressive improvements in the analysis techniques 
allowed to increase the sensitivity of this search. 
Figure \ref{fig:dms_ave} gives the details of the different $\Delta m_s$ analyses.
The combined result of LEP/SLD/CDF analyses \cite{osciWG} is shown in Figure \ref{fig:dms_ampli}:
\begin{eqnarray}
 \Delta {m}_s & > 14.4~{ps}^{-1}~~{at}~~95\%~~{C.L.} & \nonumber \\
 ~\rm{The ~ sensitivity ~ is~ } &  \Delta m_s = 19.2 ~{ps}^{-1} .   &
\label{eq:dmsresults}
\end{eqnarray}
The present combined limit implies that $B_s^0$ oscillate at least 30 
times faster than $B_d^0$ mesons.\\
The significance of the ``signal'' appearing around 17 ps$^{-1}$ is about 2.5 $\sigma$ and
no claim can be made of the observation of $B^0_s-\bar{B^0_s}$ oscillations.\\
The Tevatron experiments will measure soon $B^0_s-\bar{B^0_s}$ oscillations.

\section{Unitarity triangle parameters determination}

\begin{table*}[htb]
\caption { Values of the relevant quantities used in the fit of the CKM parameters.
In the third and fourth columns the Gaussian and the flat parts of the uncertainty are given, 
respectively \cite{parodi}. The values and the errors on $V_{cb}$ are taken from \cite{ref:ArtusoBarberio}
and are slightly different with respect to those given in equations 
\ref{eq:vcbinclres},\ref{eq:vcbexclres}.} 
\label{tab:inputs} 
\newcommand{\m}{\hphantom{$-$}}
\newcommand{\cc}[1]{\multicolumn{1}{c}{#1}}
\begin{center}
\begin{tabular}{@{}lllll}
\hline
 Parameter &  Value & Gaussian &  Uniform      & Ref. \\
           &        & $\sigma$ &  half-width   &      \\
\hline
    $\lambda$               & $0.2210$   &  0.0020    &         -     &  \cite{ref:ckmworkshop} \\
\hline
$\left | V_{cb} \right |$(excl.) & $ 42.1  \times 10^{-3}$  & $ 2.1 \times 10^{-3}$ 
                                 &              -           & \cite{ref:ArtusoBarberio}\\
$\left | V_{cb} \right |$(incl.) & $ 40.4  \times 10^{-3}$  & $ 0.7 \times 10^{-3}$ 
                                 & $ 0.8 \times 10^{-3}$    & \cite{ref:ArtusoBarberio}\\
 \hline
$\left | V_{ub} \right |$(excl.) & $ 32.5  \times 10^{-4}$ & $ 2.9 \times 10^{-4}$ 
                                 & $ 5.5 \times 10^{-4}$ & \cite{ref:ckmworkshop}\\
$\left | V_{ub} \right |$(incl.) & $ 40.9  \times 10^{-4}$ & $ 4.6 \times 10^{-4}$ 
                                 & $ 3.6 \times 10^{-4}$ & \cite{ref:ckmworkshop}\\
 \hline
$\Delta m_d$                      & $0.503~\mbox{ps}^{-1}$ & $0.006~\mbox{ps}^{-1}$ 
                                  & -- & \cite{osciWG}  \\
$\Delta m_s$  & $>$ 14.4 ps$^{-1}$ at 95\% C.L. & \multicolumn{2}{c}
{sensitivity 19.2 ps$^{-1}$} & \cite{osciWG}  \\
$m_t$ & $167~GeV$ & $ 5~GeV$ & -- & \cite{ref:top} \\
$f_{B_d} \sqrt{\hat B_{B_d}}$ & $235~MeV$  & $33~MeV$ &  $^{+0}_{-24}~MeV$  & \cite{ref:lellouch} \\
$\xi=\frac{ f_{B_s}\sqrt{\hat B_{B_s}}}{ f_{B_d}\sqrt{\hat B_{B_d}}}$ 
                                  & 1.18   & 0.04 & $^{+0.12}_{-0.00}$ & \cite{ref:lellouch} \\
 \hline
$\hat B_K$                    & 0.86   & 0.06 & 0.14 & \cite{ref:lellouch} \\
 \hline
         sin 2$\beta$             & 0.734   & 0.054 & - & \cite{ref:sin2b} \\ 
\hline
\end{tabular} 
\end{center}
\end{table*}

\begin{figure}[htb]
\includegraphics[width=75mm]{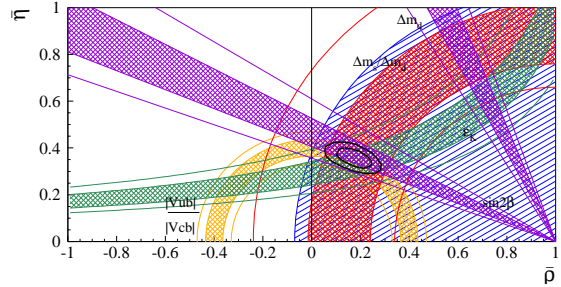}
\caption{The allowed regions for $\overline{\rho}$ and $\overline{\eta}$
(contours at 68\%, 95\%) are compared with the uncertainty bands 
for $\left | V_{ub} \right |/\left | V_{cb} \right |$, 
$\epsilon_K$, $\Delta {m_d}$,the limit on $\Delta {m_s}/\Delta {m_d} $ and sin2$\beta$.}
\label{fig:bande}
\end{figure}

Different constraints can be used to select the allowed region for the apex of the triangle 
in the $\bar{\rho}$-$\bar{\eta}$ plane. 
Five have been used sofar: $\epsilon_k$, $|V_{cb}|/|V_{cb}|$, $\Delta m_d$, the limit on 
$\Delta m_s$ and sin 2$\beta$ from the measurement of the CP asymmetry in the $J/\psi K^0$ decays.
These constraints are shown in Figure \ref{fig:bande}.

\begin{figure}[htb!]
\includegraphics[width=75mm]{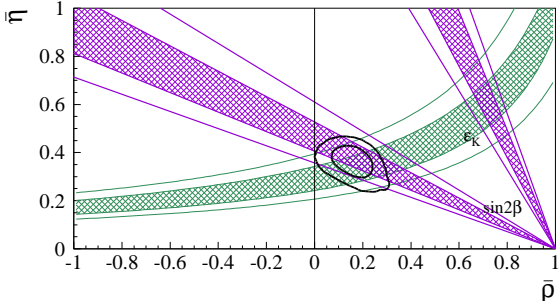}
\caption{The allowed regions for $\overline{\rho}$ and $\overline{\eta}$
(contours at 68\%, 95\%) as selected by the measurement of $\left | V_{ub} \right |/\left | V_{cb} \right |$, 
$\Delta {m_d}$, the limit on $\Delta {m_s}/\Delta {m_d} $ as compared with the bands (at 1 and 2$\sigma$)
from CP violation in the kaon ($\epsilon_K$) and in the B (sin2$\beta$) sectors.}
\label{fig:testcp}
\end{figure}

These measurements provide a set of constraints which are obtained by comparing measured 
and expected values of the corresponding quantities,  in the framework of the Standard Model (or 
of any other given model). In practice,  theoretical expressions for these constraints involve several 
additional parameters such as quark masses, decay constants of B mesons and bag-factors. The values of 
these parameters are constrained by other measurements (e.g. top quark mass) or using theoretical expectations.\\
Different statistical methods have been defined to treat the experimental and theoretical errors.
The methods essentially differ in the treatment of the latter and can be classified into two main
groups: frequentist and Bayesian. The net result is that, if the same inputs are used, the different
statistical methods select quite similar values for the different CKM parameters \cite{ref:ckmfits}. The results 
in the following are shown using the Bayesian approach.\\ 
Central values and the uncertainties taken for the relevant parameters used in these analyses are given in Table \ref{tab:inputs} \cite{parodi}.\\
The most crucial test is the comparison between the region selected by the measurements
which are sensitive only to the sides of the Unitarity Triangle and the regions selected by 
the direct measurements of the CP violation in the kaon ($\epsilon_K$) or in the B (sin2$\beta$) sector. 
This test is shown in 
Figure \ref{fig:testcp}. 

It can be translated quantitatively through the comparison between the values of 
sin2$\beta$ obtained from the measurement of the CP asymmetry in the $J/\psi K^0$ decays and the one 
determined from ``sides`` measurements:
\begin{eqnarray}
\sin 2 \beta = & 0.725^{+0.055}_{-0.065} & \rm {sides~ only}     \nonumber \\
\sin 2 \beta = & 0.734 \pm 0.054 & \rm \quad B^0 \rightarrow J/\psi K^0. 
\label{eq:sin2beta}
\end{eqnarray}
The spectacular agreement between these values shows the consistency of the Standard 
Model in describing the CP violation phenomena in terms of one single parameter $\eta$.
It is also an important test of the OPE, HQET and LQCD theories which have been used to extract the
CKM parameters.\\
Including all five constraints the results are \cite{parodi}:
\begin{eqnarray}
   \bar {\eta}  =  &  0.357 \pm 0.027              & ~(0.305-0.411)          \nonumber \\ 
   \bar {\rho}  =  &  0.173 \pm 0.046              & ~(0.076-0.260)          \nonumber \\ 
   \sin 2\beta  =  &  0.725 ^{+0.035}_{-0.031}     & ~(0.660-0.789)          \nonumber \\ 
   \sin 2\alpha =  & -0.09 \pm 0.25                & ~(-0.54-0.40)           \nonumber \\ 
   \gamma       =  &  (63.5 \pm 7.0)^{\circ}       & ~(51.0-79.0)^{\circ}    \nonumber \\ 
   \Delta m_s   =  &  (18.0^{+1.7}_{-1.5}) ps^{-1} & ~(15.4-21.7) ps^{-1}.    
\label{eq:allres}
\end{eqnarray}

The ranges within parentheses correspond to 95$\%$ probability.\\
The results on $\Delta m_s$ and $\gamma$ are predictions for those quantities 
which will be measured in  near future.

\section{Conclusions}

Many and interesting results have been presented at this conference.
Traditional main players (LEP/SLD/CLEO) are still delivering results, while the B 
factories are moving B studies into the era of precision physics.\\
Many quantities have already been measured with a good precision. The lifetimes of the B and
charm hadrons are now measured at the one/few per cent level. $|V_{cb}|$ is today known with a relative
precision better than 2$\%$. In this case, not only, the decay width has been measured, but also some of
the non-perturbative QCD parameters entering into its theoretical expression. It is a great experimental
achievement and a success of the theory description of the non-perturbative QCD phenomena in the framework 
of the OPE. Many different methods, more and more reliable, are now available 
for determining the CKM element $|V_{ub}|$. The relative precision, today, is about 10$\%$ and will be 
certainly improved in a near future at the B-factories. 
The time behaviour of $B^0-\bar{B^0}$ oscillations has been 
studied and precisely measured in the $B_d^0$ sector. The oscillation frequency $\Delta m_d$ is 
known with a precision of about 1$\%$. $B_s^0-\bar{B_s^0}$ oscillations have not been measured sofar,
but this search has pushed the experimental limit on the oscillation frequency $\Delta m_s$ 
well beyond any initial prediction. Today we know that $B_s^0$ oscillate at least 30 
times faster than $B_d^0$ mesons. The frequency of the $B_s^0-\bar{B_s^0}$ oscillations will be soon 
measured at the Tevatron. 
Nevertheless the impact of the actual limit on $\Delta m_s$ for the determination of the unitarity 
triangle parameters is crucial.\\
Many B decay branching fractions and relative CP asymmetries have been measured at the B-factories. 
The outstanding result on the determination of sin 2$\beta$ has been described in two 
plenary talks \cite{babartalk},\cite{belletalk}.
On the other hand many other exclusive hadronic B rare decays have been measured and constitute 
a gold mine for weak and hadronic physics, allowing to perform important tests of the B decay dynamics. \\
The unitarity triangle parameters are today known with a good precision. A crucial test has been already done: 
the comparison between the unitarity triangle parameters as determined with quantities sensitive to the 
sides of the unitarity triangle (semileptonic B decays and oscillations) and with the measurements of  
CP violation in the kaon ($\epsilon_K$) and in the B (sin2$\beta$) sectors. The agreement is unfortunately 
excellent. 
The Standard Model is ``Standardissimo'': it is also working in the flavour sector. 
This agreement is also an important test of the OPE, HQET and LQCD theories which have been used to extract the
CKM parameters.\\
The good news is that all these tests are at best at about 10$\%$ level. 
The current and the next facilities can surely push these tests to a 1$\%$ accuracy. It is important to note that
charm physics can play an important role in this respect (providing a laboratory for LQCD) and the Charm-factory (CLEO-C) 
will play a central role for these issues.

\section{Acknowledgements}
I would like to thank the organisers for the invitation and for having set up a very
interesting conference in a nice atmosphere.\\
I really would like to thank many colleagues from Babar, Belle, CLEO, CDF, LEP, FOCUS,
SELEX and SLD experiments, which help me in the preparation of the talk. I would also like to
remember the important work made from the members of the Heavy Flavour Working Groups who 
prepared a large fraction of the averages quoted in this note. They are all warmly thanked.\\
Finally, special thanks to E. Barberio, M. Battaglia, P. Kluit, P. Roudeau and F. Parodi both 
for the help and the support during the preparation of the talk and for the careful reading of these 
proceedings.

\end{document}